\documentclass[12pt]{emulateapj}

\shorttitle{Limits on Prompt Radio GRB Emission}
\shortauthors{Macquart}

\begin{document}

\title{On the Detectability of Prompt Coherent GRB Radio Emission}


\author{J.-P. Macquart\altaffilmark{1} 
}

\altaffiltext{1}{NRAO, P.O.\,Box 0, Socorro NM 
87801, U.S.A. (Jansky Fellow) and Astronomy Department, Mail Code 105-24, California Institute of Technology, Pasadena, CA 91125, U.S.A., {\it jpm@astro.caltech.edu}}

\begin{abstract}
Both induced Compton scattering and induced Raman scattering strongly limit the observability of the extremely bright ($\gg 10^{21}\,$K), prompt coherent radio emission recently predicted to emanate from gamma-ray bursts.  Induced Compton scattering is the main limiting factor when the region around the progenitor is not dense but when one still considers the scattering effect of the a tenuous circumburst ISM.  For a medium of density $0.01\, n_{0.01}\,$cm$^{-3}$ and a path length $L_{\rm kpc}\,$kpc and emission that is roughly isotropic in its rest frame the brightness temperature is limited to $< 3 \times 10^{18}\, \Gamma_{100}^2 \, n_{0.01}^{-1} \, L_{\rm kpc}^{-1}\,$K, where $100 \, \Gamma_{100}$ is the Lorentz factor of the frame in which the emission occurs.  Thus, for a burst at distance $D$ the predicted emission is only visible if the jet is ultra-relativistic, with $\Gamma \ga 10^3 (D/100 {\rm Mpc})$, or if the intrinsic opening angle of the emission is extremely small.  
Thus the presence or absence of such radio emission provides an excellent constraint on the Lorentz factor of the GRB outflow during the very early stages of its outburst.  Induced Raman scattering imposes an even more stringent limit independent of the emission opening angle, but only effective if GRB emission must propagate through a dense progenitor wind within $\sim 10^{15}\,$cm from the blast center.
\end{abstract}


\section{Introduction}

Several mechanisms in which extremely bright coherent radio emission may be generated within $\sim 10$\,s of the initial explosion of a gamma-ray burst have recently been proposed (Usov \& Katz 2000; Sagiv \& Waxman 2002; Moortgat \& Kuijpers 2004), and a number of strategies are being formulated to detect this radiation.  The hypothesized properties of the radiation are within the detectability of several next-generation low-frequency radio arrays, including LOFAR (Fender et al. 2006) and the dedicated GRB All-Sky Spectrometer Experiment, GASE (Morales et al. 2005).

There are a number of impediments to the detection of prompt radio emission.  Dispersion smearing by the intergalactic medium (IGM), in which the signal arrival time is strongly frequency dependent, has received considerable attention in the context of GRB radio emission (Inoue 2004).  However, this effect does not limit the detectability of impulsive emission if the signal can be dedispersed, which is possible if it is observed with sufficiently high spectral resolution.  A potentially more serious limitation arises if the IGM is highly inhomogeneous.  Multi-path propagation through such a medium results in strong temporal smearing of any impulsive signal at low frequencies (Williamson 1972; Lee \& Jokipii 1975; Cordes \& Lazio 1991).  Temporal smearing causes an irreversible loss of sensitivity when the smearing time exceeds the signal duration.  Nonetheless, temporal smearing does not currently place any solid constraints on the detectability of bursty low-frequency radio emission because little is known about the level of turbulence in the IGM on the small scales revelant to this effect.

An important property of all proposed mechanisms is that they produce radio emission with brightness temperatures in excess of $10^{21}\,$K.  In all cases the emission is generated at low frequencies, must come from a small region if the emission is prompt, and is predicted to possess a flux density in excess of $\sim 1\,$Jy.  In this paper we investigate the effect of induced Compton and induced Raman scattering as this radio emission escapes through either the dense stellar wind of the progenitor or the ISM immediately surrounding the source.  These propagation effects are potentially far more important than those previously discussed because they strongly alter the properties of radiation with high brightness temperatures $\gg 10^{10}\,$K, rendering it unobservable.  
It is pertinent to consider Raman scattering in particular because it is the dominant scattering mechanism in similar astrophysical situations involving comparably bright pulsar emission.  This mechanism is the favored explanation of eclipses in the binary pulsars PSR 1957$+$20 and PSR 1744$-$24A (Eichler 1991; Gedalin \& Eichler 1993; Thompson et al. 1994;  Melrose 1994; Luo \& Melrose 1995), for which the pulsar beams propagate through the stellar winds of their companions at certain orbital phases.



In the following section we briefly review the physics of induced Compton and Raman scattering and describe the conditions under which they are important.   In \S 3 we calculate the properties of the scattered radiation and the limits they impose on the detectability of prompt coherent radio GRB emission.  The consequences of our findings are summarized in \S4.

\section{Scattering mechanisms}


Induced Compton scattering is an important limiting process for bright sources.  It substantially enhances the contribution of spontaneous (Thomson) scattering of photons when the brightness temperature of the radiation exceeeds $\sim m_e c^2/k (\Delta \Omega)^2$, where $\Delta \Omega$ is the solid angle subtended by the radiation beam.  Induced Compton scattering limits the brightness temperature to (e.g. Wilson \& Rees 1978; Thompson et al. 1994)
\begin{eqnarray}
T_b \la \frac{m_e c^2 \, (\Delta \Omega)^2}{k \, \tau_T},
\end{eqnarray}
where $\tau_T = \int dr \sigma_T n_e(r)$ is the Thomson optical depth. If the emission is isotropic but the emitting region is approaching at a Lorentz factor $\Gamma$ one makes the replacement $\Delta \Omega^2 \rightarrow \Gamma^{-2}$ (Begelman, Ergun \& Rees 2005).

Induced Compton scatttering is important under a large range of circumstances.  It might be supposed that induced Compton scattering is less effective at wavelengths exceeding the plasma Debye length, $\lambda_D$, because the electric fields of the electrons are screened by other electrons.  However this is exactly compensated by the enhanced scattering caused by clustering of electrons around less mobile electrons (e.g.\,Thompson et al. 1994).  Nonetheless, at such long wavelengths other effects modify the scattering if the brightness temperature of the radiation is sufficiently high, and the most important of these is induced Raman scattering.

Induced Raman scattering refers to the process in which a beam of bright electromagnetic radiation propagating through a plasma induces an instability, creating turbulent Langmuir waves (plasmons) in the plasma which subsequently scatters the beam.  Raman scattering does not limit the brightness temperature in the same manner as induced Compton scattering.  When Raman scattering is effective the region in which the scattering occurs acts like a secondary photosphere, so that any beamed radiation is made isotropic, and a pulse of radiation is smeared over a timescale larger than the light travel time across the Raman photosphere (see Levinson \& Blandford 1995).  Raman scattering produces relatively small frequency shifts, $\Delta \omega/\omega \sim (kT/m_e c^2)^{1/2}$, for nonrelativistic ambient plasma temperatures, $T$, so the total radio spectrum is not strongly altered.  

Raman scattering is important provided several conditions are met.  The first is that Landau damping does not prevent propagation of the Langmuir waves.  This implies that only plasma turbulence with wavevectors $q < q_L  = 0.27 \lambda_D^{-1}$ can propagate and are available to scatter the radiation. Landau damping prevents backscattering for frequencies
\begin{eqnarray}
\nu > \nu_L = 90 n_6^{1/2} T_6^{-1/2} \,{\rm MHz}, \label{Landau}
\end{eqnarray}
and limits the maximum scattering angle to
\begin{eqnarray}
\theta_{\rm max} \sim \frac{2 \nu_L}{\nu}, \qquad \nu \gg \nu_L.
\end{eqnarray}

The second condition relates to the level at which the turbulence saturates.  This is determined by whether the plasmon occupation number saturates due to linear or nonlinear damping, giving rise to weak or strong scattering respectively.   We concentrate on the strong scattering limit applicable to the high brightness temperatures expected of coherent GRB emission.  In this regime the plasmon energy density grows exponentially with time until the production rate balances the growth rate. This occurs when the photon occupation number, $N_p = k T_B/\hbar \omega$, exceeds (Thompson et al. 1994),  
\begin{eqnarray}
N_p > N_\Gamma = \left[ \frac{3 n_e \sigma_T c \Omega}{32 \pi \Gamma_{\rm ei} }
\left( \frac{\hbar \omega^2 }{ \omega_{\rm p} m_e c^2} \right) 
\left(1+\cos^2 \theta \right) \right]^{-1}, \label{NGamma}
\end{eqnarray}
where 
$\Gamma_{\rm ei} = 32 \,n_e \,T^{-3/2}\,$s$^{-1}$ is the electron-ion collisional damping rate and $\theta$ is the angle through which the radiation is scattered.  An additional constraint is that the Langmuir turbulence growth timescale must be much shorter than the duration of the prompt emission.  The growth time depends on the ratio of the critical flux density, $S_\Gamma$, implied by eq.\,(\ref{NGamma}) to the beam flux density, $S_{\rm src}$, and is given by $\sim S_\Gamma/2 S_{\rm src} \Gamma_{\rm ei} = 0.0156 \,n_e^{-1} \,T^{3/2} (S_\Gamma/S_{\rm src})\,$s.  This growth time is much smaller than one second for bright emission that exceeds the critical flux density and conditions typical of a plasma in which induced Raman scattering is effective.


The extension of Raman scattering to magnetized plasmas with thermal effects is treated by Melrose \& Luo (1995), who show that the dominant effect of large-angle scattering is from Bernstein waves.  Bernstein waves propagate nearly perpendicular to the magnetic field near harmonics of the electron cyclotron frequency and possess sufficiently large wavenumbers to permit large-angle scattering and backscattering.   This generalization is important only in the regime in which the condition $\Omega_e \ll \omega_p$ i.e.,
\begin{eqnarray}
1.8 \times 10^4 \left( \frac{B}{1\,{\rm mG}} \right) \ll 5.6 \times 10^4 \left( \frac{n_e}{1\,{\rm cm}^{-3}} \right)^{1 \over 2},
\end{eqnarray}
is violated, otherwise the only Langmuir waves dominate the scattering.  For most cases of interest the role of the magnetic field is unimportant so the description of Raman scattering given above is sufficient; indeed, the condition $\Omega_e \ll \omega_p$ is an essential element of the Sagiv \& Waxman (2002) model.  

Brillouin scattering, the scattering of non-isotropic radiation by ion or electron acoustic waves, can also limit the radiation properties, but this is less effective than Raman scattering when the electron temperature exceeds the ion temperature (e.g. Luo \& Chian 1997).  Brillouin scattering is therefore unlikely to be important here and its effect is not considered.  However we do note that ion acoustic waves may enchance the efficiency of induced Raman scattering (Chian \& Rizzato 1994) by coupling to Langmuir waves.

In systems in which conditions for both stimulated Compton and Raman scattering are favorable the latter is the dominant process (see Thompson et al.\,1994; Levinson \& Blandford 1995).  This is because the induced Compton scattering rate is significant only if the propagation direction of the scattered photon lies inside the radio beam.  The same condition does not apply to Raman scattering.  Much larger recoil angles are possible in stimulated Raman scattering, being only limited by Landau damping or the wavenumber up to which the Langmuir turbulence extends.

Induced Compton scattering is relevant to bright emission from both AGN and pulsars (Coppi, Blandford \& Rees 1993; Begelman, Ergun \& Rees 2005; Wilson \& Rees 1978).  Induced Raman scattering is also discussed extensively in relation to eclipsing binary pulsars and constraints on coherent emission from AGN (Gedalin \& Eichler 1993; Thompson et al. 1994; Levinson \& Blandford 1995).  
The fact that extremely bright radiation is observed in pulsars implies that they must possess highly relativistic winds which mitigate the effects of scattering.
In the case of the Crab pulsar the absence of induced Compton scattering implies a wind with a minimum Lorentz factor of $\sim 10^4$ (Wilson \& Rees 1978), while induced Raman scattering imposes a similar limit of $\gg 8 \times 10^2$ (Luo \& Melrose 1994).  


\section{Limits on prompt GRB radio emission}

The limiting role played by scattering in a GRB environment depends on the properties of the medium that the bright emission must propagate through.  Two circumburst medium models which encompass the range of properties likely to be encountered are considered.  These also correspond closely to those considered for the generation of the coherent emission in the first place (cf. Sagiv \& Waxman 2002).  The first model is relevant to a burst from a system whose progenitor is a massive star, in which case the emission must propagate through the relic stellar wind.  Such winds are thought to have characteristics velocities of order $10^3\,$km\,s$^{-1}$ and mass loss rates $\dot M \sim 10^{-5}\,{\rm M}_\odot\,{\rm yr}^{-1}$ (Chevalier \& Li 1999), with temperatures $T \sim 10^6\,$K.  For an isotropic outflow the electron density at radius $r$ is 
\begin{eqnarray}
n_e &=& 3.0 \times 10^{11} \left( \frac{\dot M}{10^{-5}\,{\rm M}_\odot\,{\rm yr}^{-1}} \right) \left( \frac{v}{10^3\,{\rm km\,s}^{-1}} \right)^{-1} 
\nonumber \\ &\null& \qquad \qquad \times 
\left( \frac{r}{10^{12}\,{\rm cm}} \right)^{-2} \,{\rm cm}^{-3}.
\end{eqnarray}
The parameters of the second model are relevant to bursts resulting from mergers (e.g. NS-NS mergers), where the system has evacuated a cavity and the only available scattering material is the ionized ISM.  This plausibly has a density in the range $n_e \sim 0.001-1\,$cm$^{-3}$ and a temperature $T \sim 10^4\,$K, corresponding to that of the warm ionized medium (WIM) of our Galaxy.   The low end of the density range is appropriate to a burst that occurs in an elliptical galaxy or on the outskirts of its host galaxy, as appears to be relevant to short-duration GRBs (Prochaska, Chen \& Bloom 2006; Berger et al. 2006).

The effectiveness of the scattering mechanisms also depends on the properties of the incident radiation.
The detailed properties of the emission vary according to the radiation mechanism invoked.  Sagiv \& Waxman (2002) propose synchrotron maser emission from a purely relativistic, weakly magnetized plasma in which the plasma frequency greatly exceeds the electron cyclotron frequency.  The emission is expected to occur on a timescale of a minute after the initial gamma-ray burst at frequencies in the range $\sim 3.5$--200\,MHz, depending on the density of the medium that the blast wave expands into.  This model is expected to produce emission as strong as $S_\nu \sim 1\,$Jy for low redshift bursts (see Inoue 2004).  This corresponds to a $T_B= 10^{21-25}\,$K assuming a source size $\theta \sim \Gamma c \Delta t$.

A mechanism advanced by Usov \& Katz (2000) involves bursty synchro-Compton emission generated in the fields of low-frequency waves ahead of the GRB shock front.  This emission is expected to occur predominantly below $\sim 30\,$MHz and to last 1--$10^2$\,s.  Emission generated via this mechanism is predicted to possess a flux density $\sim 2 \times 10^6 \epsilon_B^{1.3}\,$Jy, where $10^{-4} \la \epsilon_B < 1$ is the fraction of the wind power remaining in the magnetic field at the radius at which the blast wind begins decelerating due to its interaction with ambient gas.  The associated brightness temperature is $T_B=  10^{24-29}\,$K.

Moortgat \& Kuijpers (2004) advance a more indirect mechanism for the generation of low-frequency emission from coalescing systems.  They propose that gravitational radiation from the system excites MHD wave modes at kHz frequencies which in turn undergo inverse Compton scattering off relativistic material in the GRB outflow.  Inverse Compton scattering on particles with $\gamma > 100$ boosts the radiation to observable ($\sim 100\,$MHz) frequencies.  This radiation is also predicted to be extremely bright: a source at $1\,$Gpc is expected to be visible as a $2 \times 10^6\,$Jy pulse of duration $\sim 3\,$min with a bandwidth $\sim 30\,$MHz, corresponding to $T_B \sim 10^{29}\,$K  

As all three mechanisms generate extremely bright emission, both induced Compton and Raman scattering are expected to be important limiting process outside the emission region.



\subsection{Induced Compton scattering}

We first consider the limits imposed by induced Compton scattering.  If the radiation is emitted isotropically in the rest frame of the emitting plasma the opening angle is determined only by the Lorentz factor of the frame in which the emission occurs.    For material propagating through the wind of a massive progenitor with inner radius $r_0$ the brightness temperature is constrained to be 
\begin{eqnarray}
T_B &\la& 3.0 \times 10^{22} \left( \frac{r_0}{10^8\,{\rm cm}}\right) \left( \frac{\Gamma}{100} \right)^2   \left( \frac{v}{10^3\,{\rm km\,s}^{-1}} \right) 
\nonumber \\ &\null& \qquad \qquad \times 
\left( \frac{\dot M}{10^{-5}\,{\rm M}_\odot\,{\rm yr}^{-1}} \right)^{-1}\,{\rm K}.  \label{TestIC1}
\end{eqnarray}
One can also estimate the effect of scattering associated with the propagation of the radiation through the ISM surrounding the burst.  The limit imposed by induced Compton scattering after propagation through a medium of density $n_e$ and length $L$ in the ISM is
\begin{eqnarray}
T_B \la 2.9 \times 10^{18} \left( \frac{\Gamma}{100} \right)^2 \left( \frac{n_e}{0.01\,{\rm cm}^3} \right)^{-1}
\left( \frac{L}{1\,{\rm kpc}} \right)^{-1}\,{\rm K}, \label{TestIC2}
\end{eqnarray}
which can be even more stringent than that imposed by circumburst material.
If the radiation is not isotropic in the rest frame but is instead confined to a narrow solid angle $\Delta \Omega$ the above constraints are increased by a further factor $\Delta \Omega^{-2}$.  (This is why the Galactic ISM does not constrain local pulsar brightness temperatures; their $< 100\,$nas intrinsic beam angular sizes are too small.) 

Consider the effect that these limits place on the observability of the various proposed radio emission mechanisms.  For emission of duration $\Delta t$ at a luminosity distance $D_L$ the angular size is of order $\theta \sim \Gamma \,c \, \Delta t/D_L$ so the maximum permitted flux density is
\begin{eqnarray}
S_\nu &\approx& 2.6 \, (1+z)^2 \left( \frac{T_B}{10^{20}\,{\rm K}} \right) 
\left( \frac{\nu}{30\,{\rm MHz} } \right)^2
\left( \frac{\Delta t }{100\,{\rm s}} \right)^{2} 
\left( \frac{\Gamma}{100} \right)^2 
\nonumber \\ &\null& \qquad \qquad \times 
\left( \frac{D_L}{100\,{\rm Mpc}}\right)^{-2}\quad {\rm mJy}.  \label{Slimit}
\end{eqnarray}
The above brightness temperature limits are well below those predicted for of all three proposed radiation mechanisms.  The brightness temperature constraint imposed by propagation through the surrounding ISM indicates that prompt GRB emission is so heavily suppressed by induced Compton scattering as to be unobservable unless the observing frequency exceeds $\sim 300\,$MHz or the Lorentz factor exceeds $10^3$.  

If the emission is to evade induced Compton scattering it must be intrinsically emitted into a narrow angle, $\Delta \Omega \ll 1$, in its rest frame.  For the fiducial ISM properties used in eqs.\,(\ref{TestIC1}) and (\ref{TestIC2})  induced Compton scattering is evaded if the intrinsic emission angle is less than $\sim 5 \times 10^{-4}\,(T_B/10^{25}\,K)^{-1/2}\,$ster.  Instead GRB models suggest large jet opening angles, $\sim 0.1$ (e.g. Frail et al.\,2001), suggesting that scattering is likely to be effective.  Moreover, even if the emission opening angle is narrow, the radiation may still be affected by induced Raman scattering, which scatters emission from narrowly beamed sources effectively.



\subsection{Induced Raman scattering}

For the temperatures and densities characteristic of the ISM Landau damping is effective and prevents induced Raman scattering at frequencies above $\sim 1\,$MHz.  Scattering is effective, however, if the region immediately surrounding the GRB is dense, as would be the case if the medium is filled with a dense progenitor wind.  Assuming a plasma temperature $10^6\,$K, Landau damping is ineffective for the parameters characteristic of these winds and permits large-angle scattering within a radius 
\begin{eqnarray}
r_0 &\approx& 2.4 \times 10^{15} (1+z)^{-1}
\left( \frac{\dot M}{10^{-5}\,{\rm M}_\odot\,{\rm yr}^{-1} } \right)^{1 \over 2}
\left( \frac{v}{10^3\,{\rm km\,s}^{-1}} \right)^{-{1 \over 2}}
\nonumber \\ &\null& \qquad \qquad \times 
\left( \frac{\nu}{30\,{\rm MHz}} \right)^{-1} 
\,{\rm cm},
\end{eqnarray}
of the progenitor center.  Thus strong induced Raman scattering occurs if radiation is generated within $\sim 10^{15}\,$cm of the blast center and its flux density exceeds a critical {\it in situ} value $S_\Gamma$, determined by eq.\,(\ref{NGamma}), which corresponds to an observed flux density of
\begin{eqnarray}
S_{\rm crit} &=& 0.68  \, (1+z) \left( 1 + \cos^2 \theta \right)^{-1} 
\left( \frac{n_e}{10^6\,{\rm cm}^{-3}} \right)^{1 \over 2} 
\left(\frac{D_L}{20\,{\rm Mpc} } \right)^{-2}  \nonumber \\
&\null& \,\, \times 
\left(\frac{r_0}{10^{15}\,{\rm cm} } \right)^{2} 
\left( \frac{T_e}{10^6\,{\rm K}}\right)^{-{ 3 \over 2}} \left( \frac{\nu}{30\,{\rm MHz}} \right) \, {\rm mJy}. \label{SGamma}
\end{eqnarray}
Raman scattering thus limits the radiation properties effectively if the radiation is emitted within the radius $r_0$.  Material within this radius acts as a ``Raman photosphere'', in which strong scattering through $\sim \pi/2\,$ is possible.  Scattering isotropizes any initially beamed radiation over all $4\pi$ ster, and lowers the brightness temperature of the radiation as it now appears to emanate from a region of size $r_0$.  This has the advantage of rendering the prompt emission from any bursts not beamed toward an observer visible.  However, the associated disadvantage is that any temporal signature in the emission is smeared over a timescale $\ga r_0/c \sim 10\,$hours, thus destroying a key observational signature of the bright, initially short-duration, emission.  

However, induced Raman scattering may be evaded if the emission does not occur within the radius $r_0$.  This caveat is particularly important for the GW-plasma coupling mechanism proposed by Moortgat and Kuijpers (2004) where the radii over which favourable coupling could conceivable occur at large distances from the GW production site.

\section{Conclusions}

Stimulated scattering substantially impedes the propagation of the bright coherent radio emission predicted to occur within the first $10-60$\,s of gamma-ray bursts.  These effects limit the radio emission to brightness temperatures several orders of magnitude below those predicted to be intrinsic to the emission mechanisms themselves.  The limiting effects of induced Compton scattering are only evaded if the emission is intrinsically emitted into a narrow cone or if the bulk Lorentz factor of the material in which the emission occurs is extremely high, with $\Gamma \ga 10^3 (D/100\,Mpc)$.  Thus the presence or absence of such radio emission provides an excellent constraint on the maximum Lorentz factors encountered in the GRB outflow during the very early stages of its outburst.  This limit exceeds present lower limits on the Lorentz factor, $\Gamma \ga 100-400$, imposed by the apparent absence of photon-photon pair production and Compton scattering of photons off the pair-produced $e^{\pm}$  (Lithwick \& Sari 2001).





\acknowledgments
The National Radio Astronomy Observatory is a facility of the National Science Foundation operated under cooperative agreement by Associated Universities, Inc.


\clearpage


\end{document}